\documentclass[aps,prl,epsf,superscriptaddress,twocolumn,amsmath,amssymb,showpacs,floatfix]{revtex4-1}

\usepackage{amsmath}
\usepackage{amssymb}
\usepackage{graphicx}
\usepackage{epstopdf}	
\usepackage{xspace}		

\usepackage{dcolumn}		
\usepackage{bm}			
\usepackage{subfigure}	
\usepackage{hyperref}	
  \hypersetup{colorlinks = true, 	
  citecolor = blue,		
  linkcolor = red, 		
  breaklinks = true,
  backref = true			
  }

\usepackage{upgreek}

\newcommand{\eg}{{\sl e.g.}}
\newcommand{\ie}{{\sl i.e.}}
\newcommand{\ov}{V$_\textrm{O}$\xspace}
\newcommand{\oad}{O$_\textrm{ad}$\xspace}
\newcommand{\obr}{O$_\textrm{br}$\xspace}
\newcommand{\tifc}{Ti$_\textrm{5c}$\xspace}
\newcommand{\tiint}{Ti$_\textrm{int}$\xspace}

\makeatletter

\begin{document}
\title{Observation and destruction of an elusive adsorbate with STM: O$_2$/TiO$_2$(110)}

\author{Philipp Scheiber}
\affiliation{Institute of Applied Physics, Vienna University of Technology, Wiedner Hauptstrasse 8-10, 1040 Vienna, Austria}

\author{Alexander Riss}
\altaffiliation[Present address: ]{Department of Physics, University of California, Berkeley, U.S.A.}
\affiliation{Institute of Applied Physics, Vienna University of Technology, Wiedner Hauptstrasse 8-10, 1040 Vienna, Austria}

\author{Michael Schmid}
\affiliation{Institute of Applied Physics, Vienna University of Technology, Wiedner Hauptstrasse 8-10, 1040 Vienna, Austria}

\author{Peter Varga}
\affiliation{Institute of Applied Physics, Vienna University of Technology, Wiedner Hauptstrasse 8-10, 1040 Vienna, Austria}

\author{Ulrike Diebold}
\email{diebold@iap.tuwien.ac.at}
\affiliation{Institute of Applied Physics, Vienna University of Technology, Wiedner Hauptstrasse 8-10, 1040 Vienna, Austria}

\date{\today}


\begin{abstract}

When a slightly defective rutile TiO$_2$(110) surface is exposed to O$_2$ at elevated temperatures, the molecule dissociates at defects, filling O vacancies (\ov) and creating O adatoms (\oad) on {\tifc} rows. The adsorption of molecular O$_2$ at low temperatures has remained controversial. Low-Temperature Scanning Tunneling Microscopy (LT-STM) of O$_2$, dosed on TiO$_2$(110) at a sample temperature of $\approx$\,100\,K and imaged at 17\,K, shows a molecular precursor at \ov as a faint change in contrast. The adsorbed O$_2$ easily dissociates during the STM measurements, and formation of \oad's at both sides of the original \ov is observed. 

\end{abstract}



\maketitle


The adsorption of oxygen on TiO$_2$ is a fascinating topic from both an applied and fundamental point of view. Adsorbed oxygen plays a key role in photocatalysis, both as an electron scavenger and as the oxidative species. It is important in low-temperature oxidation processes in heterogeneous catalysis and its effect on conductivity is central to semiconductor-based gas-sensing. Thus, much effort has been devoted to understand this interaction in detail  \cite{Du:2009p1661, Wendt:2008p121, Kimmel:2008p1849, Petrik:2010p1827, Henderson:1999p350, Lu:1995p816, Lu:1995p817, Zhang:2010p1619, Zhang:2010p1835, Zhang:2010p1645, Petrik:2009p1660, Rasmussen:2004p1857, Tilocca:2005p1861, Pillay:2006p1850, Dohnalek:2006p1859, Du:2010p1840, Papageorgiou:2010p1577, Onishi:1996p1866, LI:1999p910, Smith:2002p912}, using the rutile TiO$_2$(110) surface \cite {Diebold:2003p528, Pang:2008p147} as a model system. On a  stoichiometric TiO$_2$(110) surface O$_2$ only physisorbs and desorbs below $\approx$\,75\,K \cite {Dohnalek:2006p1859}. This is in agreement with density functional theory (DFT)-based calculations  that predict that excess charge, \ie,  O-deficient TiO$_2$, is essential for chemisorption \cite{Rasmussen:2004p1857, Deskins:2010p1643, Du:2010p1840}.  A slightly-reduced TiO$_2$ sample exhibits two main kinds of defects in the near-surface region: O vacancies (\ov) at two-fold coordinated, ``bridging'' atoms (\obr) and subsurface Ti interstitials (\tiint). How oxygen adsorbs at room temperature is well-understood  \cite{Du:2009p1661, Du:2010p1840, Wendt:2008p121}: the molecule dissociates at a \ov, filling the vacancy and resulting in an O adatom (\oad) that is singly-coordinated to a five-fold coordinated Ti surface atom (\tifc). Recently it has been pointed out that O$_2$ can also dissociate at subsurface \tiint's, which results in two \oad's located closely to each other at the same \tifc row \cite{Wendt:2008p121}. Oxygen exposure at slightly elevated temperature, where the \tiint's are more mobile and can migrate to the surface, results in re-growth of excess TiO$_x$ in various configurations \cite{Onishi:1996p1866, LI:1999p910, Smith:2002p912}. Scanning Tunneling Microscopy (STM) has been very helpful in unraveling these details of the interaction between O$_2$ and defects on TiO$_2$.  

What happens at low temperature is less clear, however. When O$_2$ is dosed on a cold sample, typically at a temperature around 100\,K, various desorption techniques---thermal desorption (TPD) as well as stimulated desorption by electrons (ESD) or photons (PSD)---consistently show the formation of a chemisorbed, molecular precursor \cite{Henderson:1999p350, Lu:1995p816, Lu:1995p817, Kimmel:2008p1849, Petrik:2010p1827}. The O$_2$ molecule is negatively charged, likely resembling a peroxide O$_2^{2-}$  ion, and stimulated desorption is hole-mediated \cite{Lu:1995p816, Lu:1995p817, Zhang:2010p1619, Zhang:2010p1835}. One interesting, and somewhat mysterious aspect of these experiments, however, is the fact that 
some of this molecular O$_2$ desorbs at temperatures well above room temperature, $\approx$\,400\,K
\cite{Kimmel:2008p1849, Petrik:2010p1827}. Based on TPD/ESD measurements, Kimmel and Petrik  \cite{Kimmel:2008p1849} postulated the formation of a stable O$_4^{2-}$ species (``tetraoxygen'') that survives up to these high temperatures without dissociating. To test this prediction and clarify the situation, STM measurements could be quite useful.
So far, however, molecular oxygen on TiO$_2$ was never directly observed with STM. 

Here we report low-temperature (17\,K) STM results of slightly-reduced TiO$_2$(110) surfaces exposed to O$_2$
at 100\,K. We provide evidence that O$_2$ is indeed located at \ov's, and that it is visible in STM as a rather faint change of the contrast. The O$_2$ molecule is dissociated by the STM during the measurements even at the smallest tunneling currents applied (4\,pA). 
We were also able to observe the intermediates of  O$_2$ dissociation on TiO$_2$(110). 


The experiments were performed on two different rutile TiO$_2$(110) crystals, one from CrysTec and the other one from MTI Corp. Both were cleaned by repeated cycles of sputtering (2\,keV Ar$^+$, fluence of $4 \times 10^{16}$ Ar$^+$ ions per cm$^2$) and annealing (at 1123 and 923\,K, respectively), resulting in a surface {\ov} concentration of $\approx$ 0.17\,ML.
Constant current STM measurements have been performed in a two-chamber Omicron UHV system with a base pressure below $2 \times 10^{-11}$ mbar, at sample temperatures of 17\,K (liquid He cooled) or 78\,K (liquid nitrogen cooled). Temperatures of less than 17\,K resulted in unstable tunneling. Positive sample bias voltages between 1.3 and 2.4\,V were used, and tunneling currents were varied between 0.004 and 0.4\,nA. Oxygen dosing was done by backfilling the preparation chamber with a pressure of $1 \times 10^{-9}$ mbar.


\begin{figure}[t!]
  \centering
    \scalebox{0.72}{\includegraphics{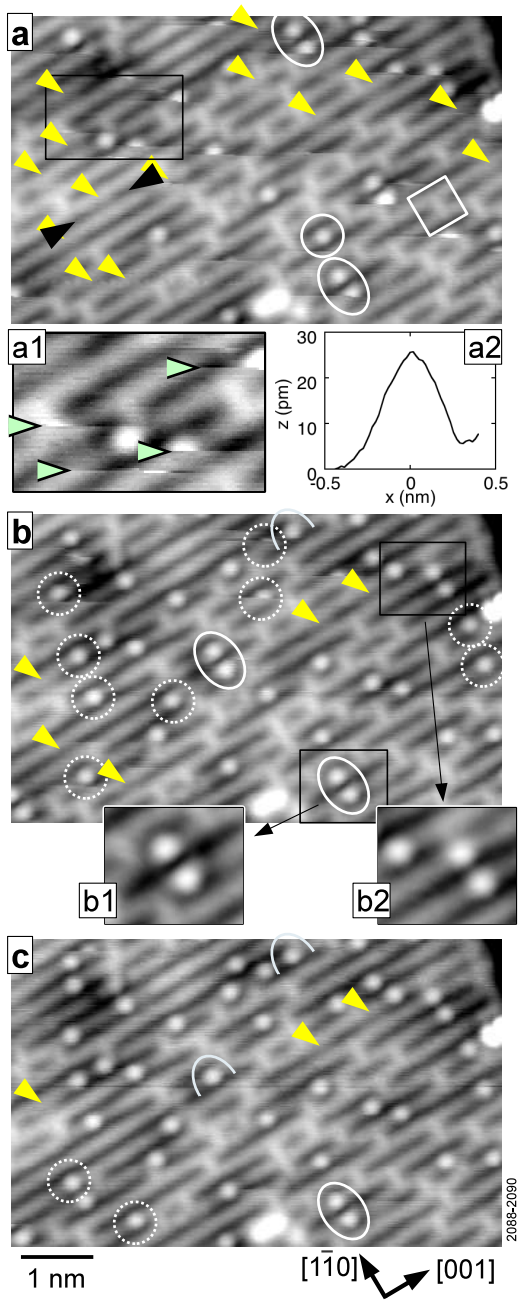}}
    \caption{(Color online) Successive STM images ($V_\textrm{sample}$ = + 1.8\,V,  $I$ = 0.05\,nA, $T_\textrm{sample}$ = 17\,K, 120 seconds per image) of a reduced rutile TiO$_2$(110) surface after exposure to 0.045\,L {O$_2$} at $\approx$\,100\,K. White box: oxygen vacancy (\ov). Circles: O adatoms ({\oad}). Dotted circles are {\oad}'s formed from {O$_2$} in the previous image. Yellow (bright) arrows: adsorbed {O$_2$}.  Ovals: two bean-shaped {\oad}'s from dissociation of {O$_2$}. a1: streakiness due to {\oad} creation (arrows); a2: averaged line profile across adsorbed {O$_2$} recorded along \obr rows (one marked by black arrows). b1, b2: comparison of \oad pairs with and without \ov in between.}
  \label{Fig1} 
  \end{figure} 

A series of STM images after dosing 0.045 Langmuirs (L; 1\,L = $10^{-6}$ torr s) O$_2$ at 100\,K is shown in Fig.\ \ref{Fig1}. It is well-established that the bright and dark lines in empty-states STM images correspond to the rows of {\tifc} and bridging oxygen (\obr)
atoms, respectively \cite{Diebold:1996p914}, and that \ov's appear as short bright lines that connect two bright {\tifc} lines. In  Fig.\ \ref{Fig1}a, 58 \ov's are visible; one is marked with a white box. Reference measurements before O$_2$ adsorption showed that an image of that size should contain 120 $\pm$ 20 \ov's.
In addition to \ov's, bright, round spots are visible on the {\tifc} rows that are clearly identifiable as O adatoms; one of these is marked with a circle in Fig.\ \ref{Fig1}a. Since (almost) each isolated {\oad} is representative of an O$_2$ molecule that has dissociated and quenched a {\ov} \cite{Wendt:2008p121}, the numbers of \oad's and \ov's  should add up to 120, the initial vacancy concentration (0.17\,ML). Even when counting the \oad's partly visible in the image, the actual number, 85, is clearly much less.

\begin{figure}
  \centering
     \scalebox{0.6}{\includegraphics{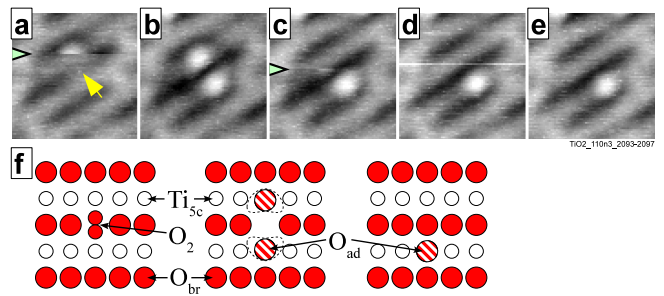}}     \caption{(Color online) Successive STM images ($V_\textrm{sample}$ = + 1.8\,V,  $I$ = 0.03\,nA, $T_\textrm{sample}$ = 17\,K) of a reduced rutile TiO$_2$ surface after exposure to 0.045\,L O$_2$ at 100\,K. Black arrows point at scan lines where an O$_2$ molecule is converted into two {\oad}'s and one of the {\oad}'s disappears by filling a vacancy. Panel (f) schematically shows the species involved.}
  \label{Fig2} 
  \end{figure}

The STM image in Fig.\ \ref{Fig1}a is streaky in a few places. This is seen more clearly in the zoom-in (a1), taken at the position of the black box. The streaks---emphasized by arrows in panel a1 of Fig.\ \ref{Fig1}---are associated with the creation of additional \oad's, which appear during scanning.
Indeed, the number of \oad's increases in consecutive STM scans (Fig.\ \ref{Fig1}b,c). Further inspection of Fig.\ \ref{Fig1}a shows an additional feature: at some positions, two neighboring bright {\tifc} rows appear smeared out and the {\obr} row between them appears slightly brighter than usual [yellow (bright) arrows].
Fig.\ \ref{Fig2}a also shows such an area; it seems that an additional, but very faint species sits on the dark {\obr} rows.
The apparent height of these features is about $\approx$\,25\,pm as shown by the line profile (Fig.\ \ref{Fig1}, panel a2), averaged over a few such sites such as the one marked by the black arrows in Fig.\ \ref{Fig1}a. For comparison, the {\tifc} rows appear 55\,pm higher than the {\obr} rows.

We will now show that these smeared out features are indicative of an O$_2$ at a \ov. In Fig.\ \ref{Fig1} they are marked with yellow (bright) arrows. Their number decreases during consecutive scans,  while the number of \oad's increases. There is a clear correlation between the position of the faint O$_2$ features and the freshly formed \oad's (dotted circles), each new {\oad} is located at a {\tifc} site next to an O$_2$ in a previous image.  

The STM images in Figs.\ \ref{Fig1} and \ref{Fig2} show yet another new feature:  Pairs of bright spots, located at adjacent {\tifc} rows. In Fig.\ \ref{Fig1}, these are marked by ovals. Such pairs of bean-shaped adatoms, which form occasionally when we dose the sample to O$_2$ at 100\,K, have not been reported before. (Note that the pairs of \oad's \cite{Wendt:2008p121} that form when O$_2$ reacts with a {\tiint} at room temperature are located at the same {\tifc} row.) Fig.\ \ref{Fig2} shows the creation of such a new adatom pair and its destruction during scanning with the STM. 
In Fig.\ \ref{Fig2}a, the smeared out feature (arrow) corresponding to an O$_2$ adsorbed at a vacancy site is still present while the tip scans across it. [The slow scan direction is +y (up) in all STM images.] When the tip arrives at the scan line marked at the left edge of Fig.\ \ref{Fig2}a, the {\obr} row suddently appears much darker and an adatom materializes on the upper {\tifc} row. Fig.\ \ref{Fig2}b also shows a second, bean-shaped adatom at the opposite side of the original O$_2$. The upper O adatom disappears in the next frame in Fig.\ \ref{Fig2}c (arrow at the edge). Adatom mobility is negligible at 17\,K, and inspection of the further surroundings shows that this {\oad} has not jumped to another location. As soon as the first adatom disappears, the remaining one changes from the original bean shape to the ``normal'', symmetric and round shape. A few similar cases are marked by ovals and half-ovals in Fig.\ \ref{Fig1}.

DFT-based studies of O$_2$ adsorbed on TiO$_2$(110) consistently predict that an O$_2^{2-}$  preferentially sits at an \ov, and that the molecule lies flat with its axis perpendicular to the rows.  Ref.\ \cite {Tilocca:2005p1861} reports Tersoff-Hamann plots of a clean, stoichiometric surface and one with an O$_2^{2-}$ in a \ov. These two plots are remarkably similar to each other, consistent with the claim that the faint, smeared out features in our STM images are indeed indicative of an O$_2$ in such a configuration. If this flat-lying O$_2$ suddenly explodes \cite{Diebold:1998p1868}, it is conceivable that the resulting two O's will land on the {\tifc} atoms adjacent to the \ov. This is a metastable situation, however, a filled {\ov} and one {\oad} will be energetically favored. So one of the \oad's will migrate back into the now empty {\ov} and fill it up as shown in the schematics in Fig.\ \ref{Fig2}f. This process can be induced by the STM tip, as evidenced by the frequent occurrence of partially imaged bean-shaped \oad's (\eg, Fig.\ \ref{Fig1}a, bottom, and Fig.\ \ref{Fig2}c).

With the STM we can easily distinguish between an across-the-row, bean-shaped adatom pair that stems from \textit{one} O$_2$ molecule and a ``pair'' of two adatoms that have formed independently and sit at adjacent positions on neighboring Ti rows by pure coincidence. For example, the pairs pointed out in panels b1 and b2 of Fig.\ \ref{Fig1}, have a marked difference in shape and brightness.  In the first case, the two adatoms are separated by an {\ov} and in the second case by an {\obr}. This affects the apparent height (brightness) of the adatoms.  Interestingly, the image contrast of the {\ov} between the two \oad's is also altered.  The gap between the two \oad's in Fig.\ \ref{Fig2}b and panel b1 of Fig.\ \ref{Fig1}, at the location where we expect the {\ov}, is quite dark.  This is in contrast to the typical appearance of an isolated  {\ov}, which is normally observed as a bright spot on a (dark) {\obr} row.  The dark {\ov} in between the newly-formed \oad pair leads to the bean-shaped appearance alluded to above.  The STM contrast on TiO$_2$ is dominated by electronic effects, and the appearance of this configuration points towards a re-arrangement of charge in the vacancy.  Based on the extensive theoretical work of the role of excess charge in oxygen adsorption it seems also conceivable that the adatoms in these two cases---either with or without a {\ov} between them---do not have the same charge.

The experimental results in Figs.\ \ref{Fig1} and \ref{Fig2} provide evidence that adsorbed molecular O$_2$ is indeed observable with STM, but they also point to the fact that the species is very unstable, and that the STM measurement itself is the trigger for most of the dissociation that is observed. The results shown here have been taken at 17\,K. Additional STM measurements at 78\,K show essentially the same features, \ie,  O$_2$ at \ov's, single \oad's that suddenly appear during scanning, and {\oad} pairs resulting from the same O$_2$. STM-induced dissociation at 78\,K is even more facile than at 17\,K, however, and in many instances most of the dissociation ocurs during the first scan of the image. At either temperature, scanning  with ``harsh'' conditions ($V_\textrm{sample} \geq + 2.4$ V, $I \geq 0.35$ nA) dissociates all O$_2$'s within one single scan, and the density of \oad's and \ov's approximately equates the number of the original \ov's prior to O$_2$ adsorption.
The effect of the tip is rather localized: when re-scanning a slightly-shifted area with ``milder'' conditions, we find that the dissociation occurs within a range of less than $\approx$\,1\,nm from the location of the tip.  

What causes the facile dissociation of adsorbed oxygen molecules during the STM scanning?  The various mechanisms for tip-induced dynamics are discussed in ref.\ \cite{Mayne:2006p1904}.  If an antibonding orbital can be accessed by the tunneling electrons, then the rate of dissociation should scale linearly with the tunneling current $I$.  If local heating is responsible, the rate should scale as $I^n$ with a higher value of $n$.  We ran extensive tests, where we varied the tunneling current over more than an order of magnitude (0.004\,nA to 0.05\,nA), but kept all other experimental parameters constant.  We chose a tunneling voltage of 1.3\,V, the lowest value where reproducible images could be obtained.
While we found a large scatter of the adatom creation range (0.0077\,ML/scan, standard deviation 0.0035\,ML/scan), the adatom creation rate shows no correlation with the tunneling current.  This clearly rules out an electron-induced process for the oxygen dissociation.  As we observe at most a weak dependence on the tunneling current, ignoring van-der-Waals interactions, the interaction between tip and adsorbed O$_2$ must be due to the electric field, which varies with the distance, and, hence, the logarithm of the current.  As the change of the shape and composition of the STM tip is not under our control, the local field under the STM tip also changes, even when we use the same tunneling voltage, explaining the scatter of adatom creation rate observed.  Hence we propose that the field is the decisive factor in how readily an adsorbed O$_2$ dissociates.  Possibly, the STM tip pushes the O$_2$ molecule into a configuration that allows an easy dissociation.  Exactly what this configuration might be is unclear at this point, but we consider it likely that a position closer to a Ti atom in the substrate will facilitate dissociation.  
It should be emphasized, however, that the product of any tip-induced process can be observed only when it is (meta)stable on the surface.  In other words, the tip helps overcoming a barrier. We consider it likely that this process also happens spontaneously at higher temperatures, and, thus, the observed {\oad} pairs should be considered the reaction intermediate of O$_2$ dissociation at TiO$_2$(110).

Desorption measurements have shown that the dissociation probability of O$_{2}^\textrm{ad}$'s is coverage-dependent, and that small amounts of O$_2$ (the coverage regime of the STM measurements shown here) dissociate more easily than larger coverages \cite{Petrik:2010p1827}.
We have varied the O$_2$ exposure in our measurements as well, but we did not observe any new features or any significant differences in adatom creation. Very recent PSD/TPD work \cite{Petrik:2010p1827} has shown that irradiation with above-band-gap photons not only desorbs O$_2$ (through hole capture) but also dissociates O$_2$ through an electron-mediated process. In addition, photon exposure was also reported to create a ``photo-blind'', thermally stable molecular O$_2$ species \cite{Petrik:2010p1827}, which was again attributed to tetraoxygen. We have searched for such species by dosing higher amounts of O$_2$ and irradiating the sample with a UV-light emitting diode (365\,nm, $\approx$\,$10^{15}$ photons 
cm$^{-2}$\,s$^{-1}$), but, again, no evidence could be found for such a species.

To summarize, we have unequivocally observed molecular O$_2$ adsorbed at oxygen vacancies on TiO$_2$ with LT-STM. Our measurements also show the difficulty of using STM as an analytical tool for learning more about this species: the STM measurement invariably dissociates the molecule even at the smallest tunneling currents, albeit with a probability that is highly tip-dependent. While it is fascinating to directly observe this dissociation, it is a nuisance when one wants to learn more about the adsorbed O$_2$ molecule itself. The STM contrast of an O$_2$ in a {\ov} is so faint that it is easily overlooked in a somewhat noisier instrument. On the other hand, the creation probability of the O adatoms is tip-dependent, thus a mere counting of the adatoms  needs to be conducted carefully to avoid erroneous conclusions. 

Acknowledgement: This work was supported by the Austrian Science Fund (FWF).

\bibliography {bib}

\end{document}